**SIMULATING DUST LATTICE WAVES IN PLASMA CRYSTALS USING THE BOX_TREE CODE.**
K. Qiao and T. W. Hyde, Center for Astrophysics, Space Physics and Engineering Research, Baylor University, Waco, TX. 76798-7316, USA, phone: 254-710-2511 (email: Ke_Qiao@baylor.edu & Truell_Hyde@baylor.edu).

**Introduction:** Dusty plasma systems play an important role both in astrophysical environments (proto-stellar clouds and ring systems) and laboratory situations (plasma processing). In general, there is still little or no evidence of strongly coupled dusty plasmas anywhere in nature. However, over the past two decades it has been shown that dust particles immersed in plasmas and radiation collect charges with their subsequent dynamics becoming influenced - or in some cases dominated - by the local electric and magnetic fields[1], [2]. (This is in addition to the usual gravitational, radiation pressure and drag effects on the particles.) There are a number of unusual observations - like the spokes at Saturn or the dust streams escaping Jupiter - that can only be understood through recognizing dusty plasma effects.

In addition, the formation and stability mechanisms for ordered colloidal (Coulomb) crystals within a tenuous dusty plasma is of great interest in protoplanetary, protostellar, accretion disk formation, spiral galaxies and dark matter research. Ever since crystal formation within a dusty plasma was discovered experimentally at the Max Plank Institute by Morfill, et al. [3], interest in plasma crystals has increased dramatically.

Recent research has focussed on defining (both theoretically and experimentally) the different types of wave mode propagation possible within plasma crystals. This is an important topic since several of the fundamental quantities for characterizing such crystals can be obtained from an analysis of the wave propagation/dispersion. There are two primary wave modes seen experimentally in dust crystals, the dust acoustic wave (DAW) and the dust lattice wave (DLW). The dust acoustic wave mode is created by pressure gradients from the ambient plasma. It was first predicted by Rao, Shukla and Yu theoretically [4] and later observed in the laboratory [5]. The dust lattice wave is similar to lattice waves observed in a normal solid lattice and is caused by the shielded coulomb interaction between dust grains in a plasma crystal. Experimentally, Pieper and Goree[6] found dust acoustic waves in horizontally extended crystals produced in RF discharges by applying a modulated voltage to a wire positioned close to the dust grains. Dust lattice waves have been observed by A. Homann et al [7], [8] in both one and two-dimensional RF discharge plasma crystals. In their situation, the waves were created using a laser as the perturbation mechanism. Although some theoretical work has also been conducted in these areas, additional numerical simulations of both DLW's and DAW's in plasma crystals are needed.

**Computer model:** In this research, both modes are studied using a Box_Tree code. The Box_Tree code is a Barnes-Hut tree code first written by Derek Richardson[9] for planetary ring and disk studies. It was later modified by Lorin Matthews[10] to include electrostatic interactions and then used by John Vasut[11] to simulate the formation of plasma crystals. It is proving to be an effective tool for modeling systems composed of large numbers of particles. The Box_Tree code models a dusty plasma system by first dividing it into self-similar patches, where the box size is much greater than the radial mean excursions of the constituent dust grains. Boundary conditions are met using twenty-six ghost boxes. A tree code is incorporated into the Box_Tree routine to allow it to deal with both gravitational and electrostatic interactions between the particles. Since most interactions can be calculated by examining the multipole interactions of collections of particles, the code scales as $N \cdot \log N$ instead of $N^2$.

**Results:** Using Box_Tree, a crystal lattice composed of dust grains was established from a random distribution. By perturbing one line of the lattice, a dust lattice wave was excited and then analyzed to obtain the fundamental lattice parameters. Several sets of results will be presented and probable applications within planetary rings will be discussed.